\documentstyle{article}

\author{ C.Mukku\thanks{ Supported by the University Grants Commission (India) 
Research Scientist scheme}\\
Dept. of Mathematics and Statistics\\
School of Mathematics and Computer/Information Sciences\\
University of Hyderabad,Hyderabad-500 046, India.}
\begin{document}
\title{Fermion Determinants and Effective Actions}

\maketitle

\abstract{
Configuration space heat-kernel methods are used to
evaluate the determinant and hence the effective action for an SU(2) doublet of
fermions in interaction with a {\it covariantly constant} SU(2) background
field. Exact results are exhibited which are applicable to {\it any} Abelian
background on which the only restriction is that $(B^{2}-E^{2})$ and
$E\cdot B$ are constant. Such fields include the uniform field and the plane
wave field. The fermion propagator is also given in terms of gauge covariant
objects. An extension to include finite temperature effects is given and the
probability for creation of fermions from the vacuum at finite temperature in
the presence of an electric field is discussed.
}

\footnote{pacs: 11.51.-q,11.51.Tk}
\pagebreak
\section{Introduction} 
Effective actions are the one tool that allow us to
probe for quantum effects in a manner that mimics classical studies. It has
grown into an important tool since its most famous and perhaps one of the first
examples - that of the Euler-Heisenberg Lagrangian obtained by ``integrating out
''the fermions in QED.

However, the impetus to study effective actions did not come until the work of
Schwinger and DeWitt. Schwinger \cite{schwinger} in his inimitable style
constructed the effective action for QED using a powerful and elegant
technique, frequently referred to as the proper-time method. While Schwinger
worked in flat space, DeWitt \cite{DeWitt} generalized his techniques to curved
spaces and related the method to the heat equation. For the first time
Schwinger had not only rederived the Euler-Heisenberg Lagrangian but went on to
find the effective action for QED in the presence of a plane-wave background
field.

More recently, the method was revived by Shore \cite{shore} who formalized the
ideas of Brown and Duff \cite{brown} with the techniques of Schwinger to enable
the calculation of effective actions for non-Abelian gauge theories with
covariantly constant background fields. We shall have more to say about such
fields in the paper. Shore's work has also been generalized to include the
effects of finite temperature \cite{mms}. It has also been shown \cite{mukku}
how the method enables one to find the effective actions for the electroweak
theory $SU(2)_L \times U(1)_Y$. However all these have dealt with only the bosonic
sectors of the gauge theories and fermionic contributions have been neglected.
The sole exception being \cite{mukku2} where it was shown that the method could
be applied to fermions both at zero and at finite temperatures equally well.
However \cite{mukku2} dealt with fermions transforming under an adjoint
representation of SU(2) with an eye to applicability to supersymmetric
theories.

More recently, there has been a resurgence of interest in the study of quantum
fields in external backgrounds both at zero and finite temperatures
\cite{elmfors}. In this paper, we consider a doublet of fermions transforming
under the fundamental representation of SU(2) and interacting with an SU(2)
Yang-Mills background gauge field. The heat kernel method is applied to find
the exact propagator and effective action contributions coming from vacuum
polarization effects of the fermions \'{a} la Euler-Heisenberg for the SU(2) 
gauge fields satisfying a covariant constancy condition.

\section{Propagators, Effective Actions and Heat Kernels}
Given a differential
operator $D(x,y)$ , the corresponding heat equation is given by
\begin{equation}
\int dz D(x,z) {\cal G}(z,y;s)=-\frac{\partial {\cal G}}{\partial s}
\end{equation}
where the Kernel , ${\cal G}(x,y,s)$ is
required to staisfy the following boundary condition
\begin{equation}
\lim_{s\rightarrow 0_{+}}{\cal G}(x,y;s)=\delta(x,y)
\end{equation}
$\delta(x,y)$ being the
Dirac delta function. The propagator or Green's function G(x,y) for the
operator $D(x,y)$ is obtained by a simple proper time (s) integration
\begin{equation} 
G(x,y)=\int_{0}^{\infty}ds {\cal G}(x,y;s) 
\end{equation}
Of course, it is understood that the kernel ${\cal G}(x,y;s)$ and the Green's 
function $G(x,y)$ will carry all the indices that the operator $ D(x,y)$ 
carries.

Effective actions upto one loop quantum effects are related to functional
determinants of operators obtained by considering quadratic fluctuations in the
classical fields. More precisely , the one loop effective action is obtained by
a Legendre transformation of the generating functional for connected Green's
function \cite{Jackiw} The standard formula for fermions is 
\begin{equation}
\Gamma(A) =ln det D 
\end{equation} 
Where $\Gamma[A]$ is the effective action
functional and A , the background field(s). This dependence on the background
field(s) comes , of course, from the operator $D(x,y)$ which contains the
background field(s) A.

Formal manipulations on $lndetD$ can be carried out to yield 
\begin{equation}
\Gamma[A]=lndet{\cal D}=-Tr\int_{0}^{\infty}\frac{ds}{s}{\cal G}(x,y;s).
\end{equation}
This formula holds as long as the operator ${\cal D}$ has no zero eigenvalues.
Traditionally, one excludes such zero modes from consideration to evaluate 
the effective action and later includes their effects through collective mode
methods.
For the purposes of this paper, we assume ${\cal D}$ has no zero eigenvalues.
Equations (3) and (5) tell us that to evaluate $G(x,y)$ and $\Gamma[A]$, all
we require is a solution to the heat equation (1).
This is where the power of the proper-time or heat kernel method is manifest.
Before we proceed to find ${\cal G}(x,y;s)$, we have to note that in equation
(5), Tr denotes a composite trace \'{a} la DeWitt; it is a trace over both 
discrete labels carried by ${\cal G}(x,y;s)$ as well as its continuous indices
;in particular,
\begin{equation}
tr_{x,y}\int_{0}^{\infty} \frac{ds}{s}{\cal G}(x,y;s)=
\int d^{n}x\int d^{n}y \delta^{n}(x,y)\int_{0}^{\infty}\frac{ds}{s}
{\cal G}(x,y;s)
\end{equation}
\section{Determinants for Dirac Operators.}
In this section, we shall examine the particular Dirac operator that concerns 
us.Notations and conventions can be found in the appendix.
The particular form of the Dirac equation we consider is
\begin{equation}
(\gamma^{\mu}D_{\mu} +m)\psi(x) =0
\end{equation}
The Dirac propagator $S(x,y)$, satisfies
\begin{equation}
(\gamma^{\mu}D_{\mu} +m)S(x,y) =\delta(x,y)
\end{equation}
Since the spinor $\psi(x)$ is in reality a doublet, $\psi_{i}(x); i=1,2$,
we must have
\begin{equation}
((\gamma^{\mu}D_{\mu})_{ij} +m\delta_{ij})S_{jk}(x,y) =\delta_{ik} \delta (x,y)
\end{equation}
having made the ``internal'' SU(2) indices explicit. For the rest of the paper,
all indices will be suppressed unless otherwise required.
The Dirac operator and the Laplacian have been studied extensively in the 
mathematical literature. The spectrum of the Laplacian has been used to get 
information on the geometrical properties of manifolds 
(``can one hear the shape of a drum?'') 
while the Dirac operator on manifolds has given topological information
through the Atiyah-Singer index theorem and its generalizations 
\cite{atiyah-singer-patodi}.
Here we shall solve our problem for the determinant of the Dirac operator by
converting it into a study of the covariant Laplacian. From equation (8) we 
can write
\begin{equation}
S(x,y) =(\gamma^{\mu}D_{\mu} +m)^{-1}\delta(x,y)
\end{equation}
implying
\begin{equation}
S(x,y) =-(\gamma^{\mu}D_{\mu} - m)[-(\gamma\cdot D)^{2} +m^{2}]^{-1}
\delta(x,y)
\end{equation}
The Green's function $G(x,y)$, for the covariant Laplacian 
\begin{equation}
{\cal D}=(-(\gamma\cdot D)^{2} +m^{2})
\end{equation}
is defined by
\begin{equation}
(-(\gamma\cdot D)^{2} + m^{2}) G(x,y) = \delta(x,y)
\end{equation}
By using the properties of the $\gamma$-matrices and
\begin{equation}
[D_{\mu},D_{\nu}]=-igF_{\mu \nu}
\end{equation}
given the $D_{\mu}=\partial_{\mu}-igA_{\mu}$, 
$F_{\mu \nu} =\partial_{\mu}A_{\nu}-\partial_{\nu}A_{\mu}
+ig[A_{\mu},A_{\nu}]$,
we find 
\begin{equation}
{\cal D}=(-(D)^{2} +\frac{1}{2}g(\sigma\cdot F) + m^{2})
\end{equation}
where 
\begin{equation}
\sigma_{\mu \nu}=\frac{i}{2} [\gamma_{\mu},\gamma_{\nu}]
\end{equation}
and $ \sigma\cdot F=\sigma_{\mu \nu} F_{\mu \nu}$.
Equation (13) now takes the following form:
\begin{equation}
(-(D)^{2} +\frac{1}{2}g(\sigma\cdot F) + m^{2})G(x,y)=\delta(x,y)
\end{equation}
The associated heat equation may be written down as 
\begin{equation}
(-(D)^{2} +\frac{1}{2}g(\sigma\cdot F) + m^{2}){\cal G}(x,y;s)
=-\frac{\partial}{\partial s}{\cal G}(x,y;s)
\end{equation}
From equations (3) and (5), it is clear that a solution of (18) will give us 
both the effective action as well as the Green's function and hence the Dirac
propagator through equation (10).
While it is easy to exponentiate the constant mass term in (18), the 
``magnetic moment'' term $(\sigma\cdot F) $ , requires a different approach.
A perturbative solution is possible through the trace of the heat kernel.
The coefficients of the series are known to be invariants of the underlying 
manifold. In our case, these invariants include invariants of the gauge field.
This particular method has been used extensively for studying gravitational 
interactions \cite{DeWitt,Barvinsky etc}.
Following the ideas of \cite{duff}and \cite{shore}, we impose a condition
on the background field. Since we do not wish to break the gauge symmetry of
the background field, we shall impose a gauge covariant condition:
\begin{equation}
D_{\mu}F_{\nu \rho}=0
\end{equation}
For an Abelian gauge field, this condition would imply a constancy of the 
background field. However, for non-Abelian gauge fields, (19) implies that only
an Abelian component survives being gauged away to zero through a gauge 
transformation \cite{shore}. Equations(18) and(19) admit the following solution:
\begin{eqnarray}
{\cal G}(x,y;s)&=&\frac{s^{-n/2}}{(4\pi)^{n/2}} \Phi(x,y)\\ \nonumber
&&\times Exp\left\{-\frac{1}{4}(x-y)_{\mu}(gFcot(gsF))_{\mu \nu}(x-y)_{\nu}\right 
\}\\    \nonumber
&&\times Exp\left \{-\frac{1}{2}trln[(gsF)^{-1}Sin(gsF)]\right \}\\ \nonumber
&&\times Exp\left \{-\frac{1}{2}gs(\sigma\cdot F) - m^{2}s\right \}
\end{eqnarray}
where
\begin{equation}
\Phi(x,y)=Pexp\left\{ ig\int_{x}^{y} A_{\mu}(z) dz_{\mu}\right \}
\end{equation}
is a path dependent phase factor with the line integral being taken over the
straight line path from x to y. It ensures the correct gauge properties for 
the kernel ${\cal G}(x,y;s)$.
From (5) we see that $\Gamma[A]$ is then given by:
\begin{eqnarray}
\Gamma[A]&=&-\frac{1}{2}Tr\int_{0}^{\infty}\frac{ds}{s}
\frac{s^{-n/2}}{(4\pi)^{n/2}} \Phi(x,y)\\ \nonumber
&&\times Exp\left \{-\frac{1}{4}(x-y)_{\mu}(gFcot(gsF))_{\mu \nu}(x-y)_{\nu}\right \}\\ \nonumber
&&\times Exp\left \{-\frac{1}{2}trln[(gsF)^{-1}Sin(gsF)]\right \}\\ \nonumber
&&\times Exp\left \{-\frac{1}{2}gs(\sigma\cdot F) - m^{2}s\right \}
\end{eqnarray}
From equation (20) it is clear that
\begin{equation}
\lim_{s\rightarrow 0}{\cal G}(x,y;s)=\lim_{s\rightarrow 0} \frac{s^{-n/2}}{(4\pi)
^{n/2}}
Exp\left \{-\frac{1}{4s} (x-y)^{2}\right \}
\end{equation}
which shows that indeed ${\cal G}(x,y;s)$ satisfies the boundary condition
as required since the expression on the RHS in (23) is just the Gaussian
representation of the Dirac delta function. It is also clear that the
asymptotic series one usually adopts for the heat kernel when dealing with
arbitrary background fields has been ``summed up'' \cite{mukku3}. Such a
summing up has been made possible by the covariant constancy condition given in
(19) implying that all derivative terms in the asymptotic series are put to
zero.
In the next section we shall consider the trace over Dirac indices in (22).
\section{Dirac Traces.}
Equation (22) has three sets of indices that need to be traced over; they are,
Lorentz, internal gauge group and Dirac spinor indices. As far as the spinor
indices are concerned, only one term contains them. The calculation is much
easier if we carry out the traces over the spinor indices first.
Let us write
\begin{equation}
D(s)=tr_{D} exp\left \{-\frac{1}{2}g s(\sigma \cdot F)\right \}
\end{equation}
where $tr_{D}$ denotes a trace over the Dirac spinor indices.
Noting that
\begin{equation}
\frac{1}{2}\left \{\sigma_{\mu \nu},\sigma_{\alpha \beta}\right \}=
\delta_{\mu \alpha} \delta_{\nu \beta}-\delta_{\mu \beta} \delta_{\nu \alpha}
+ i \epsilon_{\mu \nu \alpha \beta} \gamma_{5}
\end{equation}
where $\left \{,\right \}$ denotes the anti-commutator.
We  can therefore write
\begin{equation}
(\frac{1}{2} \sigma \cdot F)^{2}_{ij}=\frac{1}{2}(F^2)_{ij}+\frac{1}{2}
\gamma_{5}(F^{*}F)_{ij}
\end{equation}
where $F^{2}=F_{\mu \nu}F_{\mu \nu}$; $F^{*}F=\frac{i}{2}F_{\mu
\nu}\epsilon_{\mu \nu \alpha \beta}F_{\alpha \beta}$.
Since $F_{\mu \nu}$ is a non-Abelian gauge field, making the group indices
explicit, we have $F_{\mu \nu}=F^{a}_{\mu \nu} T^{a}$
where $(T^{a})_{ij}$ are the fundamental representation matrices of SU(2)
satisfying
\begin{equation}
[T^{a},T^{b}]=i\epsilon^{abc}T^{c}
\end{equation}
and we choose $T^{a}=\frac{1}{2}\sigma^{a}$ where $\sigma^{a}, a=1,2,3 $, are
the Pauli matrices (see appendix).
Equation(26) reduces to
\begin{equation}
(\frac{1}{2} \sigma \cdot F)^{2}_{ij}=\frac{1}{2}\left \{(F^{a}_{\mu \nu}
F^{b}_{\mu \nu})+\gamma_{5}(F^{a}_{\mu \nu}{*}F^{b}_{\mu \nu})\right \}\frac{1}
{4}
(\sigma^{a} \sigma^{b})_{ij}
\end{equation}
Using obvious symmetry properties of this expression, we have 
\begin{equation}
(\frac{1}{2} \sigma \cdot F)^{2}_{ij}=-(\underline{F}^{2} +\gamma_{5}
\overline{F}^{2})_{\mu \mu} \delta_{ij}
\end{equation}
Where the following gauge invariant quantities 
\begin{equation}
(\underline{F}^{2})_{\mu \nu} =\frac{1}{4} (F^{a}_{\mu \sigma}
F^{b}_{\sigma \nu})
\end{equation}
and
\begin{equation}
(\overline{F}^{2})_{\mu \nu} =\frac{1}{4} (F^{a}_{\mu \sigma}
{*}F^{b}_{\sigma \nu})
\end{equation}
have been defined.
Finally, since the eigenvalues of $\gamma_{5}$ are $\pm i$, we can write the
eigenvalues of the operator $(\frac{1}{2}\sigma \cdot F)$ as 
\begin{equation}
(\frac{1}{2} \sigma \cdot F)'_{ij}=\pm i{tr(\underline{F}^{2} \pm i
 \overline{F}^{2})}^{1/2} \delta_{ij}
\end{equation}
and
\begin{eqnarray}
D(s)&=&tr_{D}exp\left \{-\frac{1}{2}gs\sigma \cdot F\right \}\\  \nonumber
&=&\delta_{ij} 4 \Re Cosh\left \{igs[tr(\underline{F}^{2} \pm i
\overline{F}^{2})]^{1/2}\right \}
\end{eqnarray}
where $\Re $ stands for ``real part of''.
This completes the purpose of this section--that of evaluating the trace over
the Dirac spinor indices.
In the next section we shall go on to consider the other two sets of discrete
indices--Lorentz and group.
\section{Lorentz and group index traces.}
To carry out traces over the Lorentz and group indices, we notice first that
objects such as $F^{a}_{\mu \nu}$ carry both Lorentz and group indices. A
method has thus to be devised allowing us to disentangle these two sets of
indices. This is in general made possible through the definition of projection
operators.Two projection operators have already been introduced for precisely 
such a purpose, for the group that we are dealing with:SU(2).
The first is \cite{shore}
\begin{equation}
R^{ij}_{\mu \nu}=(\delta^{ij}\delta_{\mu \nu} -(\underline{F}^{2})^{-1}_{\mu
\sigma} F^{i}_{\sigma \lambda} F^{j}_{\lambda \nu})
\end{equation}
The second is \cite{mukku2}
\begin{equation}
Q^{ij}{}_{\mu \nu}=(\delta^{ij}\delta_{\mu \nu} -(\underline{F}^{2}\pm 
i\overline{F}^{2})^{-1}{}_{\mu \sigma} (F^{i}_{\sigma \lambda}
F^{j}_{\lambda \nu}\pm i {*}F^{i}_{\sigma \lambda} F^{j}_{\lambda \nu}))
\end{equation}
It is easy to verify that both of these are projection operators:
\begin{equation}
R^{ij}_{\mu \sigma} R^{jk}_{\sigma \nu} =R^{ik}_{\mu \nu}
\end{equation}
and
\begin{equation}
Q^{ij}_{\mu \sigma} Q^{jk}_{\sigma \nu} =Q^{ik}_{\mu \nu}
\end{equation}
where two simply consequences of the covariant constancy condition are
utilized:
\begin{equation}
[F_{\mu \nu},F_{\alpha \beta}]=0=[F_{\mu \nu},{*}F_{\alpha \beta}]
\end{equation}
However, these two projection operators rely on the fact that the
representation matrices are adjoint representation matrices.
We have to consider what happens when the matrices are SU(2) fundamental
representation matrices.
Since $D_{\mu}F_{\alpha \beta}=0$ implies that $[F_{\mu \nu},F_{\alpha
\beta}]=0$,
we have that 
\begin{equation}
F^{a}_{\mu \nu} F^{b}_{\alpha \beta}=F^{b}_{\alpha \beta} F^{a}_{\mu \nu},
\end{equation}
and therefore, consider
\begin{equation}
(F^{2})_{\mu \nu}{}_{ij}=F^{a}_{\mu \lambda}T^{a}_{ik}F^{b}_{\lambda \nu}
T^{b}_{kj}
\end{equation}
Since $T^{a}_{ij}$ are SU(2) fundamental representation matrices,
\begin{equation}
T^{a}_{ij}=\frac{1}{2} \sigma^{a}_{ij}
\end{equation}
it is easy to see that
\begin{equation}
(F^{2})_{\mu \nu}{}_{ij}=\frac{1}{4}F^{a}_{\mu \lambda}F^{b}_{\lambda
\nu}\delta_{ij} +\frac{1}{8}F^{a}_{\mu \lambda}F^{b}_{\lambda \nu}
[\sigma^{a},\sigma^{b}]_{ij}
\end{equation}
From equation (39) we see that the second term on the RHS is zero. Therefore we
write
\begin{equation}
(F^{2})_{\mu \nu}{}_{ij} = (\underline{F}^2)_{\mu \nu} \delta_{ij}
\end{equation}
we notice that the group indices have been separated rather trivially from the
Lorentz indices.
There is clearly no need to define any projection operator here. We may, for 
convenience sake write 
\begin{equation}
(F^{2})_{\mu \nu}{}_{ij} = (\underline{F}^2)_{\mu \sigma}P^{ij}_{\sigma \nu}
\end{equation}
where 
\begin{equation}
P^{ij}_{\mu \nu}=\delta_{ij} \delta_{\mu \nu}
\end{equation}
and clearly, it follows that
\begin{equation}
(F^{2n})_{\mu \nu}{}_{ij} = ((\underline{F}^{2})^{n}_{\mu \sigma})
P^{ij}_{\sigma \nu}; n\ge 1.
\end{equation}
The required simplifications now follow:
\begin{eqnarray}
&Exp\left \{-\frac{1}{2}trln[(gsF)^{-1}Sin(gsF)]\right \}_{ij}=&\\ \nonumber
&&Exp\left \{-\frac{1}{2}trln[(gs{\bf F})^{-1}Sin(gs{\bf F})]\right \} 
\delta_{ij}
\end{eqnarray}
and
\begin{equation}
(gFcotgsF)^{\mu \nu}_{ij}=(g{\bf F} cot(gs{\bf F}))^{\mu \nu})\delta_{ij}
\end{equation}
Therefore,
\begin{eqnarray}
&Exp\left \{-\frac{1}{4}(x-y)_{\mu}(gFcot(gsF))^{ij}_{\mu \nu}(x-y)_{\nu}
\right \}=&\\  \nonumber
&\delta_{ij}Exp\left \{-\frac{1}{4}(x-y)_{\mu}(g{\bf F}
cot(gs{\bf F}))_{\mu \nu}(x-y)_{\nu}\right \}&
\end{eqnarray}
where ${\bf F}$ is the square root of the matrix $\underline{F}^{2}$.
We should note here that this particular exponent;
$(gF cot(gsF))$ appears multiplied by factors of $(x-y)$ implying that its
contribution to the effective action or the propagator vanishes upon tracing 
over the $x$,$y$ indices. However, at finite temperature, its contribution 
would be non-zero and the result in equations (48)\& (49) would be useful then.
This completes our discussion on the traces over the Lorentz and group indices.
In the next section we shall make explicit, the effective action $\Gamma[A]$.
\section{Effective Actions, etc.}
We have seen that the effective action for our SU(2) fundamental fermions in
interaction with a covariantly constant SU(2) gauge field can be written as
\begin{eqnarray}
\Gamma[A]&=&-\frac{1}{2}Tr\int_{0}^{\infty}\frac{ds}{s}\frac{s^{-n/2}}
{(4\pi)^{n/2}}
\Phi(x,y)\\ \nonumber
&&\times Exp\left \{-\frac{1}{4}(x-y)_{\mu}(gFcot(gsF))_{\mu \nu}(x-y)_{\nu}
\right \}\\ \nonumber
&&\times Exp\left \{-\frac{1}{2}trln[(gsF)^{-1}Sin(gsF)]\right \}\\ \nonumber
&&\times Exp\left \{-\frac{1}{2}gs(\sigma \cdot F) - m^{2}s\right \}
\end{eqnarray}
Using the results of the last two sections on Dirac, Lorentz and group traces
along with equation(6), we find
\begin{eqnarray}
\Gamma[A]&=&-\frac{1}{2}tr_{group}\int d^{n}x \int_{0}^{\infty}\frac{ds}{s}
\frac{s^{-n/2}}{(4\pi)^{n/2}}\\ \nonumber
&& \Re Cosh\left \{igs[tr(\underline{F}^{2}+i\overline{F}^{2})]^{1/2}\right \}
\\ \nonumber
&&\times Exp\left \{-\frac{1}{2}trln[(gsF)^{-1}Sin(gsF)]\right \} 
Exp\left \{- m^{2}s\right \}
\end{eqnarray}
Before proceeding, we must remember that any quantum calculation involves the
appearance of infinities and therefore a process of regularization and 
renormalization has to be carried out.
We know that the proper-time method is an ``invariant regularization'' method
\cite{schwinger} and so we need only worry about subtracting the infinities in
(51).
It is well known that in the proper-time integration, the traditional UV 
(UltraViolet) infinities arise in the limit of $s\rightarrow 0$ while IR
(InfraRed) infinities arise at the upper limit of the integration: $s\rightarrow
\infty$. I.e; the short distance behaviour is given by $s\rightarrow 0$ and the
behaviour at large scales is given by $s\rightarrow \infty$.
In this paper, we shall not worry about the infinities or their structure.
They have been well studied in the literature and while they are useful for
the study of renormalization properties, we are more interested in the finite 
structure of the effective actions.
To this end, we simply extract the infinities in the proper-time integration
of (51) and subtract the terms from the integrand, leaving a finite expression 
for $\Gamma[A]$.
Note that the mass term in the exponent in (51) tells us that there will be
no divergent terms in the integrand for $s\rightarrow \infty$ hence no IR
divergences.
While in four dimensions ($n=4$), in the limit of $s\rightarrow 0$, divergences
will arise from terms which are at most quadratic in s. Therefore, expanding the 
integrand in powers of s and subtracting terms upto $s^{2}$, we find a finite
$\Gamma[A]$;
We shall carry out these subtractions at the end.
Using (47) we can write (51) in a form where the group indices are explicit:
\begin{eqnarray}
\Gamma[A]&=&-\frac{1}{2}tr_{group}\int_{0}^{\infty}\frac{ds}{s}\int d^{n}x
\frac{s^{-n/2}}{(4\pi)^{n/2}}\\ \nonumber
&&4\Re Cosh\left \{igs[tr(\underline{F}^{2}+i\overline{F}^{2})]^{1/2}\right \} 
\delta_{ij}\\ \nonumber
&&\times Exp\left \{-\frac{1}{2}trln[(gs{\bf F})^{-1}Sin(gs{\bf F})]\right \} 
exp(-m^{2}s)
\end{eqnarray}
Therefore,
\begin{eqnarray}
\Gamma[A]&=&-\int_{0}^{\infty}\frac{ds}{s}\int d^{n}x
\frac{s^{-n/2}}{(4\pi)^{n/2}}\\ \nonumber
&&4\Re Cosh\left \{igs[tr(\underline{F}^{2}+
i\overline{F}^{2})]^{1/2}\right \}\\
&&\times Exp\left \{-\frac{1}{2}trln[(gs{\bf F})^{-1}
Sin(gs{\bf F})]\right \} exp(-m^{2}s)
\end{eqnarray}
Now, we require to find the eigenvalues of $\underline{F}^{2}$ in order to
express $\Gamma[A]$ in terms of the two Abelian gauge invariants
\begin{equation}
{\cal F}_{1}=\frac{1}{4}F^{a}_{\mu \nu}F^{a}_{\mu \nu}
\end{equation}
and
\begin{equation}
{\cal F}_{2}=\frac{1}{4}F^{a}_{\mu \nu}{*}F^{a}_{\mu \nu}
\end{equation}
Since an Abelian group admits only these two invariants, the fact that our 
non-Abelian field satisfying the covariant constancy condition (19) and hence
(38) tells us that we shall also have only these two invariant Lorentz scalars 
appearing.
The eigenvalues of $\underline{F}^{2}$ (and hence of $\overline{F}^{2}$) are
\cite{shore}
\begin{equation}
f^{\pm}=-{\cal F}_{1} \pm \sqrt{{\cal F}_{1}^{2}+{\cal F}_{2}^{2}}
\end{equation}
with each eigenvalue occuring with a degeneracy factor of two.
It follows then that:
\begin{equation}
Exp\left \{-\frac{1}{2}trln[(gs{\bf F})^{-1}Sin(gs{\bf F})]\right \}=
\frac{g^{2}s^{2}\sqrt{f^{+}}\sqrt{f^{-}}}{Sin(gs\sqrt{f^{+}})Sin(gs\sqrt{f^{-}})}
\end{equation}
where 
\begin{equation}
\sqrt{f^{\pm}}=\frac{i}{\sqrt{2}}[({\cal F}_{1}+i{\cal F}_{2})^{1/2}
\pm ({\cal F}_{1}-i{\cal F}_{2})^{1/2}]
\end{equation}
or,
\begin{equation}
Exp\left \{-\frac{1}{2}trln[(gs{\bf F})^{-1}Sin(gs{\bf F})]\right \}=
\frac{g^{2}s^{2}{\cal F}_{2}}{\Im Cosh(gsX)}
\end{equation}
with
\begin{equation}
X^{2}=2({\cal F}_{1}+i{\cal F}_{2})
\end{equation}
and $\Im$ denoting ``imaginary part of''.\\
Noting that $tr\underline{F}^{2}=-{\cal F}_{1}$, and
$igs\sqrt{[tr(\underline{F}^{2}+i\overline{F}^{2})]}=-gsX$, we can finally write
\begin{eqnarray}
\Gamma[A]&=&-\frac{1}{2}\int d^{n}x \int_{0}^{\infty}\frac{s^{-n/2}}{(4\pi)^{n/2}}
\\ \nonumber
&&4\left \{\Re Cosh(gsX)(\frac{2 g^{2}s^{2}{\cal F}_{2}}{\Im Cosh(gsX)})\right \}
Exp(-m^{2}s)
\end{eqnarray}
In writing the final expression, we have cheated just a little. Because we have 
the dual of $F_{\mu \nu}$,$ {*}F_{\mu \nu}$ appearing in $\Gamma[A]$, the 
spacetime dimension must be four so that ${*}F$ is also a two form along with 
$F$. Also, as early as equation (25), because of the appearance of $\gamma_{5}$ 
and $\epsilon_{\mu \nu \alpha \beta}$, we should have put $n=4$.
The reason we have done so at that stage is that for the case of  a purely 
magnetic background field, ${\cal F}_{2} =0$ and so $\Gamma[A]$ can live in 
spacetime dimensions $\ge 4$--after the Dirac traces are done $\!$.
However, notice that we do not really need to go to $n$ dimensions (usually
done for regularizing purposes) because regularization has been effected by
the proper-time itself as long as the proper-time integration is carried out
last \cite{schwinger}.
For zero-field, the fermions are free and therefore, the effective action must
go to zero. Thus, in four dimensions,
\begin{equation}
\Gamma[A]=-\frac{1}{4\pi^{2}}\int_{0}^{\infty}\int d^{4}x \left \{g^{2}s^{2}
{\cal F}_{2} (\frac{\Re Cosh(gsX)}{\Im Cosh(gsX)}) -1\right \}exp(-m^{2}s)
\end{equation}
Comparing this with the expression obtained by Schwinger (equation $(3\cdot 44)$ 
of \cite{schwinger}) shows the changes brought about by going to a non-Abelian 
group, SU(2) and its fundamental representation, from the QED result. One may 
say that equation (62) is for SU(2) QCD, what Schwinger's result is for U(1) QED
and we see that for covariantly constant fields, the contribution from the doublet
is simply twice the QED result $\!$. In some sense, a decoupling has taken place.
To conclude this section, let us consider the UV infinities of $\Gamma[A]$:
Expanding the integrand around $s=0$ and retaining terms that lead to divergences,
we find the following:
\begin{equation}
-\frac{1}{6\pi^{2}}\int_{0}^{\infty}g^{2}\frac{ds}{s}{\cal F}_{1}e^(-m^{2}s).
\end{equation}
Adding the classical Lagrangian to the one-loop effective Lagrangian, we write
the final result in the following form:
\begin{eqnarray}
\Gamma[A]&=&-\left \{1+\frac{g^{2}}{6\pi^{2}}\int_{0}^{\infty}\frac{ds}{s}
e^{-m^{2}s}\right \}{\cal F}_{1}\\ \nonumber
&&-\frac{1}{4\pi^{2}}\int_{0}^{\infty}\frac{ds}{s^{3}}e^{-m^{2}s}\left \{
g^{2}s^{2}{\cal F}_{2}(\frac{\Re Cosh(gsX)}{\Im Cosh(gsX)}) -1-\frac{2}{3}g^{2}
s^{2}{\cal F}_{1}\right \}
\end{eqnarray}
In the next section we shall find the propagators for the fermions using our
results for the heat kernel before proceeding to consider finite temperature
effects.
\section{Propagators.}
In this section we shall exhibit the propagator for our SU(2) doublet fermions 
using the relationship between the Green's function for the Laplacian and the
propagator exemplified by equations (3),(10) and (11).
\begin{equation}
S(x,y)=-(\gamma_{\mu}D_{\mu} -m)\int_{0}^{\infty} ds {\cal G}(x,y;s)
\end{equation}
Since $D_{\mu}F_{\alpha \beta}=0$, it is easy to show \cite{shore} that $D_{\mu}
\Phi(x,y)$ is given by
\begin{eqnarray}
D_{\mu}\Phi(x,y)&=&\frac{i}{2}F_{\mu \lambda}(x)\Phi(x,y)(x-y)_{\lambda}
\\ \nonumber
&=&\frac{i}{2}\Phi(x,y)F_{\mu \lambda}(y)(x-y)_{\lambda}
\end{eqnarray}
Therefore the propagator is simply
\begin{eqnarray}
S(x,y)&=&\int_{0}^{\infty}\frac{s^{-n/2}}{4\pi)^{n/2}}[m-\frac{i}{2}\gamma_{\mu}
F_{\mu \nu}(x)(x-y)_{\nu}\\ \nonumber
&&-\frac{1}{2}(gFcotgFs)_{\mu \nu}(x-y)_{\nu}]\Phi(x,y)\\ \nonumber
&&\times Exp\left \{-\frac{1}{4}(x-y)_{\mu}(gFcotgsF)_{\mu \nu}(x-y)_{\nu}
\right \}\\ \nonumber
&&\times Exp\left \{-\frac{1}{2}trln[(gsF)^{-1}Sin(gsF)]\right \}\\ \nonumber
&&\times Exp\left \{-\frac{1}{2}g(\sigma \cdot F)s -m^{2}s\right \}
\end{eqnarray}
Note that this is the {\underline exact} propagator for SU(2) fundamental fermio
ns in an external SU(2) gauge field satisfying the covariant constancy condition 
(19). In addition, since we have seen that there is a simple separation of the 
group and Lorentz indices, $S(x,y)$ can be reduced to:
\begin{eqnarray}
S(x,y)&=&\int_{0}^{\infty}\frac{s^{-n/2}}{4\pi)^{n/2}}[m-\frac{i}{2}\gamma_{\mu}
F_{\mu \nu}(x)(x-y)_{\nu}\\ \nonumber
&&-\frac{1}{2}(gFcotgFs)_{\mu \nu}(x-y)_{\nu}]\Phi(x,y) \frac{g^{2}s^{2}
{\cal F}_{2}}{\Im Cosh(gsX)}\\ \nonumber
&&\times Exp\left \{-\frac{1}{4}(x-y)_{\mu}(g{\bf F}cotgs{\bf F})_{\mu \nu}
(x-y)_{\nu}\right \}\\ \nonumber
&&\times Exp\left \{-\frac{1}{2}g(\sigma \cdot F)s -m^{2}s\right \}
\end{eqnarray}
This is as much of a simplification as one can achieve without further 
specialization of the background fields (such as to a purely magnetic field).
In the next section, we shall extend our considerations to include finite
temperature effects using the imaginary time formalism.
\section{Finite temperature effects.}
After the pioneering work on finite temperature effects in gauge theories
\cite{dolanjackiwbernardweinberg}, there has been a resurgence of interest in 
the subject in recent years. On the one hand, non-equibilibrium phenomena are
being tackled while equilibrium thermodynamics is being applied (through both
the imaginary and real time formalisms) to the study of QED and QCD plasma
formations. This has become very important as the progress in the study of
heavy-ion collisions holds promise of new developments leading to a better 
understanding of the structure of matter in extreme environments.

In this paper, we shall apply the imaginary time formalism as modified for 
applicability through the heat kernel \cite{mukku2,mms}, to find the finite
temperature corrections to the effective action.
In configuration space, the heat kernel method can be generalized to include 
finite temperature effects through the method of images and the finite temperature
heat kernel, ${\cal G}^{\beta}(x,y;s)$ is constructed from the zero temperature
kernel, ${\cal G}(x,y;s)$ as follows \cite{mms}:
\begin{equation}
{\cal G}^{\beta}(x,y;s)=\sum_{p=-\infty}^{\infty} (-1)^{p}{\cal G}(x-p\lambda 
\beta,y;s)
\end{equation}
wher $(-1)^{p}$ ensures the correct boundary conditions for the fermions,
$\beta=1/kT$, with $k$ being Boltzmann's constant and $\lambda$ is a unit vector
in the time direction.
The generalization of the $lndet$ is straighforward and is given by:
\begin{eqnarray}
lndet_{\beta}{\cal D}&=&-Tr\int_{0}^{\infty}\frac{ds}{s}{\cal G}^{\beta}(x,y;s)
\\ \nonumber
&=&-Tr\sum_{p=-\infty}^{\infty}(-1)^{p}\int_{0}^{\infty}\frac{ds}{s}{\cal G}
(x-p\lambda \beta,y;s)
\end{eqnarray}
while the Green's function is given as:
\begin{eqnarray}
G^{\beta}(x,y)&=&\int_{0}^{\infty}ds {\cal G}^{\beta}(x,y;s)\\ \nonumber
&=&\int_{0}^{\infty}ds {\cal G}(x-p\lambda \beta,y;s)
\end{eqnarray}
From equation (20) it is easy to see that
\begin{eqnarray}
{\cal G}^{\beta}(x,y;s)&=&\frac{s^{-n/2}}{(4\pi)^{n/2}}\sum_{p=-\infty}^{\infty}
\Phi(x-p\lambda \beta,y)\\ \nonumber
&&\times Exp\left \{ -\frac{p^{2}\beta^{2}}{4}\lambda_{\mu}(gFcot(gsF))_{\mu \nu}
\lambda_{\nu}\right \}\\ \nonumber
&&\times Exp\left \{ -\frac{1}{2} trln[(gsF)^{-1}Sin(gsF)]\right \}\\ \nonumber
&& \times Exp\left \{ -\frac{1}{2}gs(\sigma \cdot F) -m^{2}s\right \}
\end{eqnarray}
In this expression, the phase factor $\Phi(x-p\lambda \beta,y)$ is the only 
term that needs to be dealt with a little care as it leads to non-equilibrium
situations along with a loss of gauge covariance. For the special case of purely
magnetic fields however, it reduces to unity.
At this stage, for simplicity, we shall impose a second condition on the backgr
ound field by requiring
\begin{equation}
\Phi (x-p\lambda \beta,y)=1
\end{equation}
A brief discussion of the consequences of such a restriction has been given
in \cite{mms} while a more detailed analysis may be found in 
\cite{grosspisarskiyaffe} where they argue (for the case of QCD) that periodic 
configurations for which this condition does not hold, contribute negligibly 
to the effective action. For our purposes, it is sufficient to assume this 
additional condition holds and we note that the only term that is affected by 
the finite temperature corrections is the term containing $(gFcot(gsF))$, the 
term which drops out of the zero temperature effective action.
In particular, the Dirac spinor index traces will be identical to the zero
temperature case.
Thus, as the finite temperature effective action, $\Gamma^{\beta}[A]$ is
given by
\begin{eqnarray}
\Gamma^{\beta}[A]&=&-\frac{1}{2}Tr\int_{0}^{\infty}\frac{ds}{s}\frac{s^{-n/2}}
{(4\pi)^{n/2}}\\ \nonumber
&&\times \sum_{p=-\infty}^{\infty}Exp\left \{ -\frac{p^{2}\beta^{2}}{4}
\lambda_{\mu}(gFcot(gsF))_{\mu \nu}\lambda_{\nu}\right \}\\ \nonumber
&&\times Exp\left \{ -\frac{1}{2} trln[(gsF)^{-1}Sin(gsF)]\right \}\\ \nonumber
&& \times Exp\left \{ -\frac{1}{2}gs(\sigma \cdot F) -m^{2}s\right \},
\end{eqnarray}
we can carry out the Dirac traces and simplifying the expression, we find
\begin{eqnarray}
\Gamma^{\beta}[A]&=&-\frac{1}{2}Tr\int_{0}^{\infty}\frac{ds}{s}\frac{s^{-n/2}}
{(4\pi)^{n/2}}\\ \nonumber
&&\times \sum_{p=-\infty}^{\infty}4 \Re Cosh\left \{igs[tr(\underline{F}^{2}+i
\overline{F}^{2})]^{1/2}\right \}\\ \nonumber
&&\times Exp\left \{ -\frac{1}{2} trln[(gsF)^{-1}Sin(gsF)] -m^{2}s\right \}.
\end{eqnarray}
For the group index traces, we see from equation (48) that
\begin{eqnarray}
[\lambda_{\mu}(gFcot(gsF))_{\mu \nu}\lambda_{\nu}]^{ij}&=&\delta^{ij}
[\lambda_{\mu}(g{\bf F}Cot(gs{\bf F}))_{\mu \nu}\lambda_{\nu}] \\ \nonumber
&=&\delta^{ij}(g{\bf F}cot(gs{\bf F}))_{00}
\end{eqnarray}
Hence, 
\begin{eqnarray}
\Gamma^{\beta}[A]&=&-\frac{1}{4\pi^{2}}\int_{0}^{\infty}\frac{ds}{s^{3}}\int 
d^{4}x \\ \nonumber
&&\sum_{p=-\infty}^{\infty}Exp\left \{-\frac{p^{2}\beta^{2}}{4}(g{\bf F}
Cot(gs{\bf F}))_{00}\right \}\\ \nonumber
&&\times (\Re Cosh\left \{igs[tr(\underline{F}^{2}
+i\overline{F}^{2})]^{1/2}\right \})\left 
\{ \frac{g^{2}s^{2}\sqrt{f^{+}}\sqrt{f^{-}}}{Sin(gs\sqrt{f^{+}})
Sin(gs\sqrt{f^{-}})}\right \} e^{-m^{2}s}
\end{eqnarray}
The evaluation of $(g\sqrt{\underline{F}^{2}}Cot(gs\sqrt{\underline{F}^{2}}%
)_{00}$ can be carried out either by constructing the diagonalizing matrix
or by noting that $F$,$F^{2}$ and $(g\sqrt{\underline{F}^{2}}Cot(gs\sqrt{%
\underline{F}^{2}})_{00}$ are all diagonalized by the same diagonalizing
matrix. Simple matrix algebra then gives: 
\begin{eqnarray}
(g{\bf F}Cot(gs{\bf F}))_{00}&=&(g\sqrt{\underline{F}^{2}}
Cot(gs\sqrt{\underline{F}^{2}})_{00} \\ \nonumber
&=&E^{2}\frac{(Cot(gs\sqrt{f^{-}})- Cot(gs\sqrt{f^{+}})}
{(\sqrt{f^{-}}-\sqrt{f^{+}})}
\end{eqnarray}
where we notice that the expected loss of covariance has taken place with an
appearance of the electric field.
Unfortunately, due to the appearance of singular structures, it is not possible
to deduce the result in the case of a purely electric or purely magnetic field
from that given in equation (78) for the general case.
We can return to the expression in equation (49) along with the result from 
(58) to find that for either a pure magnetic or pure electric field, we can 
write
\begin{equation}
(g{\bf F}Cot(gs{\bf F}))_{00}= g\sqrt{f^{+}} Cot(gs\sqrt{f^{+}})
\end{equation}
and of course, in the limit of zero field, this reduces to $(1/s)$ as it should.
Let us conclude this section by examining the singular structure of the 
integrand in $\Gamma^{\beta}[A]$ for the case of a purely electric background 
field. For the case of a purely electric background field, the integrand in 
the effective action, $\Gamma^{\beta}[A]$ has a term $ gECot(gsE)$ as in the case 
of \cite{schwinger}.
The singularities therefore lie at propertimes $s_{n}=n\pi/gE$. Normally (as
at zero temperature), this would lead to an imaginary contribution to 
$\Gamma[A]$.
But for non-zero temperatures, we notice that we have an exponent with a 
non-zero contribution. Therefore, we have the
interesting result that for a purely electric background field, the imaginary
part of the effective action gets an additional exponential factor. 
There are some fermionic systems in 2+1 dimensions which seem to show signals of
a phase transition from zero to non-zero temperatures\cite{mermin}.
A more detailed analysis of the exponent in equation (78) with its temperature
dependence is needed for a better understanding of the behaviour of the 
doublet in the presence of a general covariantly constant background field.
Further work on this aspect of our paper is under progress.
For the purposes of this paper, this completes the study of finite 
temperature corrections. In the next section we shall present our conclusions 
along with prospects for future work and associated applications.

\section{Conclusions.}
In this paper we have shown how the Euclidean configuration space heat kernel
can be used to find the effective Lagrangian for SU(2) doublet fermions 
interacting with a covariantly constant SU(2) Yang-Mills field. The result 
is exact and mimics that of Schwinger. Since both a uniform field and a 
plane-wave field satisfy the covariant constancy condition, our result holds for 
both these background configurations.
We have also been able to write down the exact propagator in such backgrounds
by virtue of having a closed form solution of the heat equation for finite 
temperature through the
imaginary time formalism and the method of images, has yielded a result which 
suggests a phase transition.
For a purely electric background field, we have a situation where the 
probability for pair creation from the vacuum while being non-zero at zero 
temperature, vanishes for non-zero temperatures.
This warrants further investigation and a more detailed study is in progress.
The most important future work we envisage is to utilize the SU(2) propagators
to study SU(2) QCD plasma formations.
Since much experimental work is being focussed on high energy collisions and
the quark-gluon plasma.
Such studies would be most relevant.
Moreover, from a gauge theoretic point of view, the calculations presented in 
this paper point towards a more complete analysis of effective actions for 
unified gauge theories such as $SU(2)_{L}\times U(1)_{Y}$.

\newpage
\section{References.}
\bibitem[1]{schwinger} J.Schwinger, {\it On gauge invariance and vacuum 
polarization},\\
\indent Phys. Rev.{\bf 82}(1951) 664.
\bibitem[2]{DeWitt} B.S.DeWitt, {\it Dynamical Theory of Groups and Fields}, 
Gordon and Breach (1964).
\bibitem[3]{shore}G.Shore, Annals of Physics (NY) {\bf 137}(1981)262.
\bibitem[4]{brown}M.R.Brown and M.J.Duff, Phys.Rev.{\bf D11}(1975)2124.
\bibitem[5]{mms}J.A.Magpantay, C.Mukku and W.A.Sayed, Annals of Physics (NY)
{\bf 145}(1983)27.
\bibitem[6]{mukku}C.Mukku, Annals of Physics (NY) {\bf 162}(1984)335.
\bibitem[7]{mukku2}C.Mukku, ICTP Trieste, preprint IC/82/37.
\bibitem[8]{elmfors}Per Elmfors, {\it Dispersion Relations from the Hard thermal 
loop}\\
\indent {\it Effective Action in a Magnetic Field}\\
\indent preprint, CERN-TH/95-274 (hep-ph/9501314) and references therein.\\
\indent K.W. Mak, Phys.Rev. {\bf D49}(1994)6939.
\bibitem[9]{Jackiw}R.Jackiw, Phys.Rev.{\bf D9}(1974)1686.
\bibitem[10]{atiyah-singer-patodi}M.F.Atiyah, R.Bott, V.K.Patodi, Inv. Math.{\bf 
19}(1973)279.
\bibitem[11]{Barvinsky etc}G.A.Vilkovisky,{\it Heat Kernel: Recontre Entre 
Physiciens et Mathematiciens}\\
\indent preprint, CERN-TH6392/92. Published in Proceedings,\\ 
\indent Institut de Recherche Math\'{e}matique Avanc\'{e}e,\\
\indent Universit\'{e} Louis Pasteur, Strasbourg.
\bibitem[12]{duff}M.J.Duff and M.Ram\'{o}n-Medrano, Phys.Rev.{\bf D12}(1975)3357.
\bibitem[13]{mukku3}C.Mukku, Phys.Rev.{\bf D45}(1992)2916.
\bibitem[14]{dolanjackiwbernardweinberg}C.Bernard, Phys.Rev. {\bf D9}(1974)3312
.\\
\indent L.Dolan and R.Jackiw, Phys.Rev.{\bf D9}(1974)3320.\\
\indent S.Weinberg, Phys.Rev.{\bf D9}(1974)3357.
\bibitem[15]{grosspisarskiyaffe}D.J.Gross, R.D.Pisarski and L.G.Yaffe, Rev.Mod
.Phys.{\bf 53}(1981)43.
\bibitem[16]{mermin}N.D.Mermin and H.Wagner, Phys. Rev.Lett. {\bf 17}(1966)1133.
\newpage

\newcommand{\seqnoll}{\setcounter{equation}{0}}
\renewcommand{\theequation}{\thesection.\arabic{equation}}

\seqnoll
\appendix
\section{Appendix}

In this appendix, we give a set of consistent notations and conventions 
which have been used in the text.
The Dirac matrix algebra satisfies:
\begin{equation}
\left \{\gamma_{\mu},\gamma_{\nu}\right \}=2\delta_{\mu \nu}; \hspace{0.5in} \mu 
\nu=0,1,2,3.
\end{equation}
braces signify an anticommutator: $\{ a,b\}=ab+ba$.
While,
\begin{equation}
\gamma_{\mu}^{2}=1; \hspace{0.5in} \gamma_{\mu}^{\dag}=\gamma_{\mu}; \hspace{0.5in} \mu=0,1,2,3.
\end{equation}
and
\begin{equation}
\gamma_{i}=\left( \begin{array}{cc}
0 & \sigma_{i}\\ 
\sigma_{i} & 0 \end{array} \right); \hspace{0.5in}i=1,2,3.
\end{equation}
where
\begin{equation}
\sigma_{1}=\left( \begin{array}{cc} 0 & 1\\ 1 & 0 \end{array} \right); 
\sigma_{2}=\left( \begin{array}{cc} 0 & -i\\ i & 0 \end{array} \right);
\sigma_{3}=\left( \begin{array}{cc} 1 & 0\\ 0 & -1 \end{array} \right)
\end{equation}
are the Pauli matrices.
We also define
\begin{equation}
\gamma_{0}=\left( \begin{array}{cc} I_{2} & 0\\ 0 & -I_{2} \end{array} \right); 
I_{2}=\left( \begin{array}{cc} 1 & 0\\ 0 & 1 \end{array}\right)
\end{equation}
and
\begin{equation}
\gamma_{5}=i\gamma_{0}\gamma_{1}\gamma_{2}\gamma_{3}; \hspace{0.5in} 
\gamma_{5}^{\dag}=-\gamma_{5};\hspace{0.5in} \gamma_{5}^{2}=-1.
\end{equation}
This implies that $\gamma_{5}$ has $(\pm i)$ as eigenvalues with multiplicity 
two.
Defining
\begin{equation}
\sigma_{\mu \nu}=\frac{i}{2}\left[ \gamma_{\mu},\gamma_{\nu}\right],
\end{equation}
it is easy to establish that
\begin{equation}
\frac{1}{2}\left \{\sigma_{\mu \nu},\sigma_{\alpha \beta}\right \}=
\delta_{\mu \alpha} \delta_{\nu \beta}-\delta_{\mu \beta} \delta_{\nu \alpha}
+ i \epsilon_{\mu \nu \alpha \beta} \gamma_{5}.
\end{equation}
Lastly, the SU(2) algebra is chosen to satisfy the following Lie bracket:
\begin{equation}
\left[ T^{a},T^{b}\right]=i\epsilon^{abc}T^{c}; \hspace{0.5in} a,b,c=1,2,3.
\end{equation}
While the generators of the fundamental representation are chosen to be:
\begin{equation}
T^{a}=\frac{1}{2}\sigma^{a}, \hspace{0.5in} a=1,2,3.
\end{equation}

\end{document}